\documentclass[article]{aastex631}
\usepackage{etoolbox}
\usepackage{xpatch}

\makeatletter
\xpatchcmd{\endthebibliography}{\newpage}{}{}{}
\makeatother

\usepackage{blindtext}
\usepackage{amsmath}
\usepackage{amssymb}
\usepackage{comment}
\usepackage{dsfont}
\usepackage{bbm}
\usepackage{bm}
\usepackage{soul}
\usepackage[table]{xcolor}

\usepackage{dsfont}

\shorttitle{}
\shortauthors{}
\begin{document}

\title{Near Resonance, Less Ordered: Radius and Mass Ordering in Compact Multi-planet Systems}

\author[0009-0004-1149-9887]{Agustin Heron} 
\affiliation{Department of Astronomy, Indiana University, Bloomington, IN 47405, USA.}

\author[0000-0002-7846-6981]{Songhu Wang}
\affiliation{Department of Astronomy, Indiana University, Bloomington, IN 47405, USA.}

\author[0000-0003-0412-9314]{Cristobal Petrovich}
\affiliation{Department of Astronomy, Indiana University, Bloomington, IN 47405, USA.}

\begin{abstract}
Compact multi-planet systems exhibit a tendency for planetary radii to follow ordered sequences with orbital position, suggesting that these non-random patterns preserve a fossil record of their formation. Here, we investigate whether ordering of planet radii and masses with orbital period is related to proximity to mean-motion resonances. We compare near-resonant and non-resonant systems using complementary rank- and inversion-based measures of ordering, while accounting for differences in planet multiplicity and uncertainties in the measured planetary properties. We find that radii and masses are less monotonically ordered in near-resonant systems than in non-resonant systems, with the primary radius analysis providing stronger evidence than the mass analysis. This result is opposite to the naive expectation that dynamical instabilities should erase the primordial ordering of initially resonant systems and leave non-resonant systems in a more disordered state.
\end{abstract} 

\keywords{Exoplanet systems --- Exoplanet dynamics --- Orbital resonances --- Exoplanet evolution}

\section{Introduction} \label{sec:intro}

Planets orbiting the same star tend to be more similar in radius and mass, and more regularly spaced, than planets drawn randomly from the broader population \citep{Weiss+2018, Wang2017, Millholland+2017, Weiss+2020, Gilbert+2020, Goyal+2022}. Similar properties can emerge during planet formation through processes such as \textit{in-situ} accretion coupled to oligarchic growth \citep{Dawson+2016, MacDonald+2020, Mishra+2021, Kambara+2025}, sequential formation at planet-induced pressure bumps \citep{Xu+2024}, and assembly within localized planetesimal rings \citep{Batygin+2023, Jiang+2023}. Moreover, arguments based on low-energy configurations suggest that nearly equal masses and regular orbital spacing can be preferred architectures \citep{Adams2019, Adams+2020}. Subsequent dynamical instabilities, including giant impacts and planet mergers, can further reshape these correlations rather than simply erase them \citep{Izidoro+2017, Goldberg+2022, Lammers+2023}. 

These results indicate that different formation and evolutionary pathways can produce \emph{peas-in-a-pod} patterns. Thus, to help distinguish among these pathways, we can examine not only how similar planets are within a system, but also the ordering of their properties with orbital period. This distinction matters because uniformity does not necessarily imply ordering: planets may have similar radii or masses without these following a preferred monotonic sequence with orbital position.

The importance of planetary ordering was highlighted by \citet{Kipping2018}, who showed that \textit{Kepler} multi-planet systems exhibit more ordered radius sequences than expected from randomized arrangements. For transiting systems, radius ordering is particularly accessible because orbital periods and planet-to-star radius ratios are among the most precisely measured observables in exoplanet characterization. Planetary masses are available for a smaller and less homogeneous subset of systems, but provide complementary information as they are less sensitive to thermal and atmospheric structural effects than radii, and directly influence gravitational interactions between planets.

As planets form and evolve, disk-driven migration can capture neighboring planets into mean-motion resonances and assemble resonant chains, provided that migration remains convergent and that eccentricity damping from the parent disk does not lead to overstable librations \citep{Goldreich+2014, Deck+2015}. In disks with smooth surface density profiles, both conditions depend on planetary masses and their ordering, suggesting that the formation and survival of resonant chains may favor monotonic arrangements of planet masses. 

Proximity to resonance has also been linked to the \emph{peas-in-a-pod} pattern. \citet{Goyal+2023} found that near-resonant systems exhibit enhanced intra-system radius uniformity compared with non-resonant systems. This difference may reflect dynamical instabilities that reshape planet similarities established during formation and migration. If the same processes also weaken planetary ordering, non-resonant systems might exhibit greater radius and mass disorder than their near-resonant counterparts. However, whether the observed connection between resonance proximity and intra-system uniformity extends to planetary radius and mass ordering remains unclear.



\section{Sample Selection} \label{sec:sample_selection}

We constructed separate samples for the radius and mass ordering analyses using confirmed planets and planet candidates listed in the NASA Exoplanet Archive \citep{Christiansen+2025} as of 2026 March 13, drawing from the Planetary Systems \citep{ps}, Kepler Objects of Interest Cumulative \citep{cumulative}, K2 Planets and Candidates \citep{k2pandc}, and TESS Project Candidates tables. The radius sample comprises transiting systems from the \textit{Kepler}, \textit{K2}, and \textit{TESS} missions, and includes both confirmed planets and planet candidates, with the inclusion of candidates supported by the low false-positive rates found in \textit{Kepler} multi-transiting systems \citep{Lissauer+2012, Rowe+2014}. The mass sample comprises systems of confirmed planets with reported masses. Despite these different observational requirements, both samples are selected using the same astrophysical cuts in orbital period, planetary radius, and planet multiplicity.

For both samples, we restrict the analysis to compact systems with $P<100$ days and exclude any system containing a planet with nominal radius $R_p>4\,R_\oplus$. This approximately limits the sample to the super-Earth and sub-Neptune regime and avoids systems containing giant planets, whose formation and dynamical evolution may differ substantially. We further require at least three planets per system, so that our metrics probe the architecture of a planetary sequence rather than the relative properties of a single planet pair. Because relatively few systems satisfy the mass requirements at this multiplicity, the mass sample is drawn from all transit surveys and is therefore less observationally homogeneous than the radius sample.

To avoid counting the same system more than once in the radius sample, we first identified entries associated with the same host star using common TIC identifiers. For \textit{Kepler} candidates, we converted KIC identifiers to TIC identifiers using the publicly available cross-identification table from the \texttt{kic2tic}\footnote{\url{https://github.com/jradavenport/kic2tic}} project. After grouping entries by host star, we compared their orbital periods to identify potential duplicate planet entries. We then resolved these duplicates and removed entries lacking measurements of orbital period or planetary radius. Finally, to avoid mixing observational selections within individual systems, we retained only systems whose selected planets were discovered by the same mission sample: \textit{Kepler}, \textit{K2} or \textit{TESS}.

After applying the selection criteria, the radius sample contains 184 \textit{Kepler} systems with 626 planets, 34 \textit{K2} systems with 115 planets, and 34 \textit{TESS} systems with 113 planets (see Table \ref{tab:catalog_information} for more details). The mass sample contains 17 systems with 51 planets.

\section{Near-Resonant Classification} \label{sec:resonance}

To identify potential resonant pairs within our sample, we use the simplest proxy for resonance: the orbital period ratio between neighboring planets. In particular, we use the fractional deviation from exact period commensurability
\begin{equation}
    \Delta = \frac{P_{\rm out}/P_{\rm in}}{p/q}-1,
\end{equation}
where $P_{\rm out}$ and $P_{\rm in}$ are the observed orbital periods of the outer and inner planets, respectively, and $p$ and $q$ are small positive integers defining a resonance of order $|p-q|$. A small value of $|\Delta|$ indicates proximity to a period commensurability, but does not by itself establish that the corresponding resonant angles are librating. We therefore use the term \emph{near-resonant} to describe pairs selected based on their orbital period ratios.

Following \citet{Dai+2024}, we classify pairs near first-order MMRs as near-resonant when
\begin{equation}
    -0.015 \leq \Delta \leq +0.03,
    \label{ec:delta_1st}
\end{equation}
and pairs near second-order MMRs when
\begin{equation}
    -0.015 \leq \Delta \leq +0.015.
    \label{ec:delta_2nd}
\end{equation}

For first-order resonances, the asymmetric interval reflects both the observed excess of planet pairs wide of exact commensurability and the tendency of first-order resonant libration centers to occur at $\Delta>0$ \citep{Batygin+2013b, Petrovich+2013, Choksi+2020}. The wider population at positive $\Delta$, together with the corresponding deficit immediately interior to resonance, constitutes the well-known peak--trough structure of the \textit{Kepler} period-ratio distribution \citep[e.g.,][]{Fabrycky+2014, Lithwick+2012}. This structure is generally interpreted as the signature of resonant interactions modified by dissipative processes such as eccentricity damping in the protoplanetary or planetesimal disk and tidal dissipation \citep{Lithwick+2012, Batygin+2013a, Chatterjee+2015, Choksi+2020}. For second-order resonances, the symmetric interval is motivated by the observed period-ratio distribution and dynamical calculations \citep{Bailey+2022, Dai+2024}.

We consider the 2:1, 3:2, 4:3, 5:4 and 6:5 first-order MMRs, as well as the 3:1, 5:3, 7:5 and 9:7 second-order MMRs, following \citet{Dai+2024}. For every system in our sample, we evaluate all adjacent planet pairs ordered by their nominal orbital periods. Orbital period uncertainties are negligible relative to the adopted percent-level intervals in $\Delta$ and are therefore not propagated in our classification. In particular, typical uncertainties in $\Delta$ are $\lesssim10^{-6}$ for \textit{Kepler} systems and $\lesssim10^{-4}$ for \textit{K2} and \textit{TESS} systems \citep{Dai+2024}. Then, a planetary system is classified as near-resonant if at least one neighboring pair satisfies the corresponding criterion above, and as non-resonant otherwise.

\begin{figure*}[t!]
    \centering
    \includegraphics[width=1\linewidth]{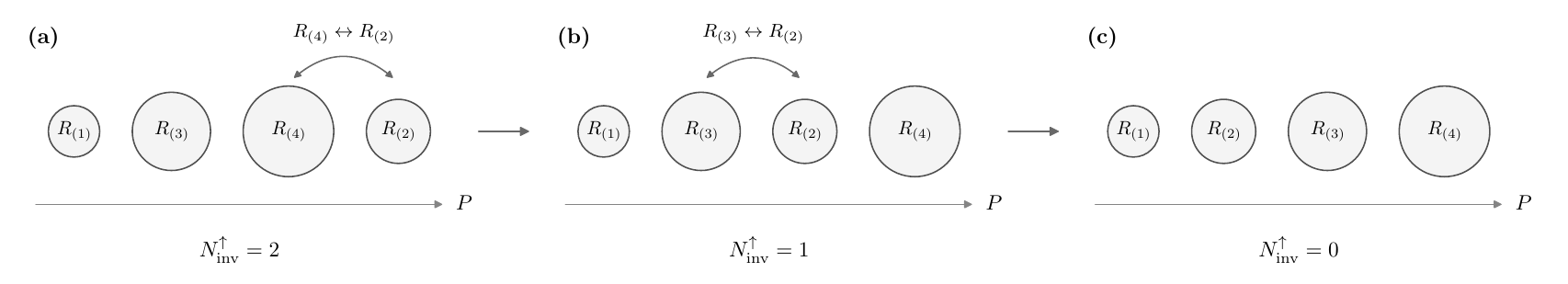}
    \caption{Number of inversions as a measure of radius ordering. Planets are arranged from left to right in order of increasing orbital period $P$, while $R_{(i)}$ denotes the $i$-th radius rank, such that $R_{(1)}<R_{(2)}<R_{(3)}<R_{(4)}$. Panel (a) shows a 4-planet configuration with two inversions relative to monotonically increasing radius order, $N_{\rm inv}^{\uparrow}=2$. Successive adjacent panels remove one inversion at a time, producing $N_{\rm inv}^{\uparrow}=1$ in panel (b) and the fully increasing configuration, $N_{\rm inv}^{\uparrow}=0$, in panel (c).\\}
    \label{fig:schematic}
\end{figure*}

\section{Planetary Ordering Metrics}
\label{sec:metrics}

Previous studies have quantified planetary ordering using entropy-based measures \citep{Kipping2018}. These statistics quantify how commonly a given ordering pattern occurs among all possible arrangements of planetary properties, however, they do not directly measure departures from monotonicity. For example, consider two five-planet systems with radius-rank sequences $(1,2,3,5,4)$ and $(2,3,4,5,1)$, ordered by increasing orbital period. In both cases, planetary radii increase through the first four planets before decreasing at the outermost planet. Thus, both sequences have identical entropy under the definition of \citet{Kipping2018}. However, the first sequence differs only slightly from a fully monotonically increasing arrangement, whereas the second places the smallest planet at the outermost position. 

Thus, to directly quantify monotonic ordering, we adopt Spearman's rank correlation coefficient, previously used to characterize planetary monotonicity by \citet{Gilbert+2020}, together with inversion-based statistics (see Figure \ref{fig:schematic}). For each planetary system, we arrange planets in order of increasing orbital period, $P_1 < P_2 < \cdots < P_N$, and consider the corresponding sequence of a planetary property, $\{W_1, W_2, \ldots, W_N\}$, where $W$ denotes either radius or mass. We then apply both metrics to the full planetary sequence, thereby characterizing the monotonic ordering of the system as a whole rather than focusing on individual planet pairs.

\subsection{Spearman's Rank Correlation Coefficient}

We use the direction-independent Spearman coefficient $D_{\rm S}$, to quantify the degree of monotonic ordering in planetary radius or mass, irrespective of whether $W$ increases or decreases with orbital position. For $N$ planets arranged by increasing orbital period, we define
\begin{equation}
    D_{\rm S} = 1 - \left| 1-\frac{6}{N(N^2-1)}\sum_{i=1}^{N}(r_i-i)^2 \right|,
\end{equation}
where $r_i$ is the rank of the planetary property $W_i$ among the $N$ planets, assuming distinct values of $W$. A value of $D_{\rm S}=0$ corresponds to either a fully monotonically increasing or decreasing sequence, whereas larger values indicate greater departures from monotonic ordering. Because $D_{\rm S}$ depends only on the relative ranks of planetary properties, it does not distinguish between small and large differences in planetary radii or masses.

\subsection{Number of Inversions} \label{sec:inversions}

As a complementary measure of monotonic ordering, we use the number of inversions, which counts how many planet pairs are out of order in radius or mass relative to a fully monotonic sequence. Unlike the Spearman coefficient $D_S$, the inversion metric allows us to introduce a fractional threshold, so that planets with sufficiently similar radii or masses are treated as effectively indistinguishable when assessing their ordering.

For planets organized by increasing orbital period, we define the increasing-direction number of inversions as
\begin{equation}
    N_{\rm inv}^{\uparrow}(\varepsilon_W)=\sum_{i=1}^{N-1}\sum_{j=i+1}^{N}\mathds{1}\left(\frac{W_i}{W_j} > 1+\varepsilon_W\right),
\label{eq:increasing_inversions}
\end{equation}
where $\mathds{1}(\cdot)$ is the indicator function and $\varepsilon_W$ is a fractional threshold in the planetary property $W$. Since the planets are organized by increasing orbital period, $i$ denotes an inner planet and $j$ an outer planet. Then, an inversion is counted when the inner planet's value of $W$ is larger than that of the outer planet by more than the specified fractional threshold $\varepsilon_W$.

Similarly, the decreasing-direction number of inversions is defined as
\begin{equation}
    N_{\rm inv}^{\downarrow}(\varepsilon_W)=\sum_{i=1}^{N-1}\sum_{j=i+1}^{N}\mathds{1}\left(\frac{W_j}{W_i} > 1+\varepsilon_W\right),
\label{eq:decreasing_inversions}
\end{equation}
which counts planet pairs that depart from a fully monotonically decreasing sequence.

For $\varepsilon_W=0$ and distinct values of $W$, these definitions reduce to the usual number of inversions of a permutation (e.g., \citealt{Knuth1998}). Then, every inner--outer planet pair contributes to exactly one of the two directional inversion counts, such that
\begin{equation}
    N_{\rm inv}^{\uparrow} + N_{\rm inv}^{\downarrow}=\binom{N}{2}=\frac{N(N-1)}{2},
\label{eq:inversion_complementarity}
\end{equation}
where each directional count is non-negative and bounded by $N(N-1)/2$. 

In the absence of a fractional threshold, the increasing- and decreasing-direction statistics are complementary: a configuration that is highly ordered in one direction is necessarily highly disordered in the other. In this limit, $N_{\rm inv}^{\uparrow}$ corresponds to the minimum number of adjacent swaps required to transform an observed sequence into a fully monotonically increasing one, as illustrated in Figure~\ref{fig:schematic}. For $\varepsilon_W>0$, planet pairs with fractional differences at or below the adopted threshold contribute to neither directional inversion count, so the complementarity relation in Equation~\ref{eq:inversion_complementarity} no longer holds.

To additionally characterize ordering without assuming a preferred direction, we define the direction-independent number of inversions as
\begin{equation}
    N^{\uparrow\downarrow}_{\rm inv}(\varepsilon_W)=\min\left[N_{\rm inv}^{\uparrow}(\varepsilon_W),N_{\rm inv}^{\downarrow}(\varepsilon_W)\right],
\label{eq:direction_independent_inversions}
\end{equation}
which is zero for both fully monotonically increasing and decreasing sequences, with larger values indicating greater departures from monotonic ordering in either direction.

For both planetary radius and mass, we compute the increasing-direction and direction-independent number of inversions. We evaluate both metrics at $\varepsilon_W = 0,\ 0.1,\ 0.2,$ and $0.3$, corresponding to fractional thresholds of $0\%$, $10\%$, $20\%$, and $30\%$, respectively.

\section{Data Analysis} \label{sec:analysis}

To compare planetary ordering between near-resonant and non-resonant systems, we account for the dependence of the ordering metrics on planet multiplicity, uncertainties in the reported radii and masses, and differences in the multiplicity distributions of the two populations. We address these effects through standardization relative to permutation distributions, Monte Carlo propagation of measurement uncertainties, and multiplicity-weighted population comparisons, respectively.

\subsection{Propagation of Measurement Uncertainties} \label{sec:property_resampling}

For each planetary system, we propagate the reported measurement uncertainties through the ordering analysis using Monte Carlo sampling. In each realization, we draw a value of $W$ for each planet from an asymmetric Gaussian distribution defined by its reported value and lower and upper uncertainties, truncated at zero to exclude nonphysical negative radii or masses. Orbital periods, planet multiplicities, and near-resonant classifications are kept fixed.

For the radius sample, the reported planetary radius uncertainties include a contribution from the host star radius, which can lead to an overestimation of the uncertainty in the relative ordering of planetary radii when sampled independently within the same system. To mitigate this effect, we subtract the stellar radius contribution in quadrature for each planet, assuming independent uncertainties in the planet-to-star radius ratio and the stellar radius
\begin{equation}
    \left(\frac{\widetilde{\sigma}_p^{\pm}}{R_p}\right)^2=\left(\frac{\sigma_p^{\pm}}{R_p}\right)^2-\left(\frac{\sigma_{\star}^{\pm}}{R_\star}\right)^2
\end{equation}
where $\sigma_p^{\pm}$ and $\widetilde{\sigma}_p^{\pm}$ denote the reported and corrected planetary radius uncertainties, respectively, while $\sigma_{\star}^{\pm}$ denotes the stellar radius uncertainties. The superscripts $\pm$ indicate upper and lower uncertainties, respectively. We apply this correction when the required stellar information is available and both corrected variances are non-negative; otherwise, we retain the reported planetary radius uncertainties $\sigma_p^{\pm}$.

We also use the Monte Carlo realizations to assess whether the direction-independent Spearman coefficient is sensitive to small differences between planetary radii or masses. Because this statistic depends only on relative ranks, we repeat the analysis using sub-samples selected according to the mean pairwise fractional contrast in $W$. For each realization, we calculate
\begin{equation}
    \bar{\delta}_{W}=\frac{2}{N(N-1)}\sum_{i=1}^{N-1}\sum_{j=i+1}^{N}\left[\frac{\max(W_i,W_j)}{\min(W_i,W_j)}-1\right],
\end{equation}
where $N$ is the system multiplicity. For a given threshold $\varepsilon_W$, we retain a system if $\bar{\delta}_{W}>\varepsilon_W$ in more than $90\%$ of its Monte Carlo realizations. Increasing $\varepsilon_W$ therefore restricts the analysis to systems with larger mean pairwise contrasts, without changing the Spearman statistic itself. System eligibility is determined once for each threshold and held fixed throughout the analysis.

\subsection{Standardization Relative to Permutation Distributions} \label{sec:standardization}

Comparing planetary ordering across systems with different numbers of planets requires accounting for the dependence of the ordering metrics on multiplicity. For the inversion metrics, the maximum possible number of inversions increases with the number of planets, making raw counts difficult to compare across multiplicities. Although the direction-independent Spearman coefficient is bounded between zero and one, its possible values also depend on multiplicity. Moreover, at low multiplicity, the small number of possible configurations produces discrete, generally non-Gaussian distributions. We therefore standardize each ordering statistic relative to a reference distribution constructed from all possible arrangements of the observed planetary properties.

For each planetary system and Monte Carlo realization, we construct the reference distributions by evaluating the ordering statistics for all $N!$ possible permutations of the sampled planetary radii or masses, where $N$ is the number of planets in the system. This procedure considers every possible arrangement of the same planets while preserving their multiplicity and sampled radii or masses. Reference distributions are constructed separately for each metric, with the inversion statistics evaluated at each adopted fractional threshold $\varepsilon_W$. We then determine the position of the ordering statistic $X$ for the sampled sequence within its corresponding reference distribution by calculating the midpoint cumulative probability:
\begin{equation}
    u=\frac{1}{N!}\left[n_{<}(X)+\frac{1}{2}n_{=}(X)\right],
\label{eq:midpoint_percentile}
\end{equation}
where $n_{<}(X)$ and $n_{=}(X)$ denote the numbers of configurations with ordering statistics smaller than and equal to the measured value $X$, respectively. Because different planetary configurations can yield the same ordering statistic, the midpoint treatment accounts for these ties by assigning them the center of their corresponding cumulative probability interval.

Finally, we transform this midpoint cumulative probability $u$ into a Gaussian-equivalent standardized score using:
\begin{equation}
    Z_X \equiv \Phi^{-1}(u) = \sqrt{2}\ \mathrm{erf}^{-1}(2u-1),
    \label{eq:normal_score}
\end{equation} 
where $\Phi^{-1}$ is the inverse cumulative distribution function of the standard normal distribution and $\mathrm{erf}^{-1}$ denotes the inverse error function. This transformation does not assume that the underlying random reference distribution is Gaussian; instead, it maps equivalent cumulative probabilities across different multiplicities onto a common scale. Thus, a value of $Z_X=0$ corresponds to the midpoint ($u=0.5$) of the reference distribution. For both direction-independent metrics, $Z_X<0$ indicates stronger monotonic ordering than expected from random ordering, whereas $Z_X>0$ indicates weaker ordering.

\subsection{Multiplicity-Weighted Population Comparison} \label{sec:population_comparison}

\begin{figure*}[t]
    \centering
    \includegraphics[width=1\linewidth]{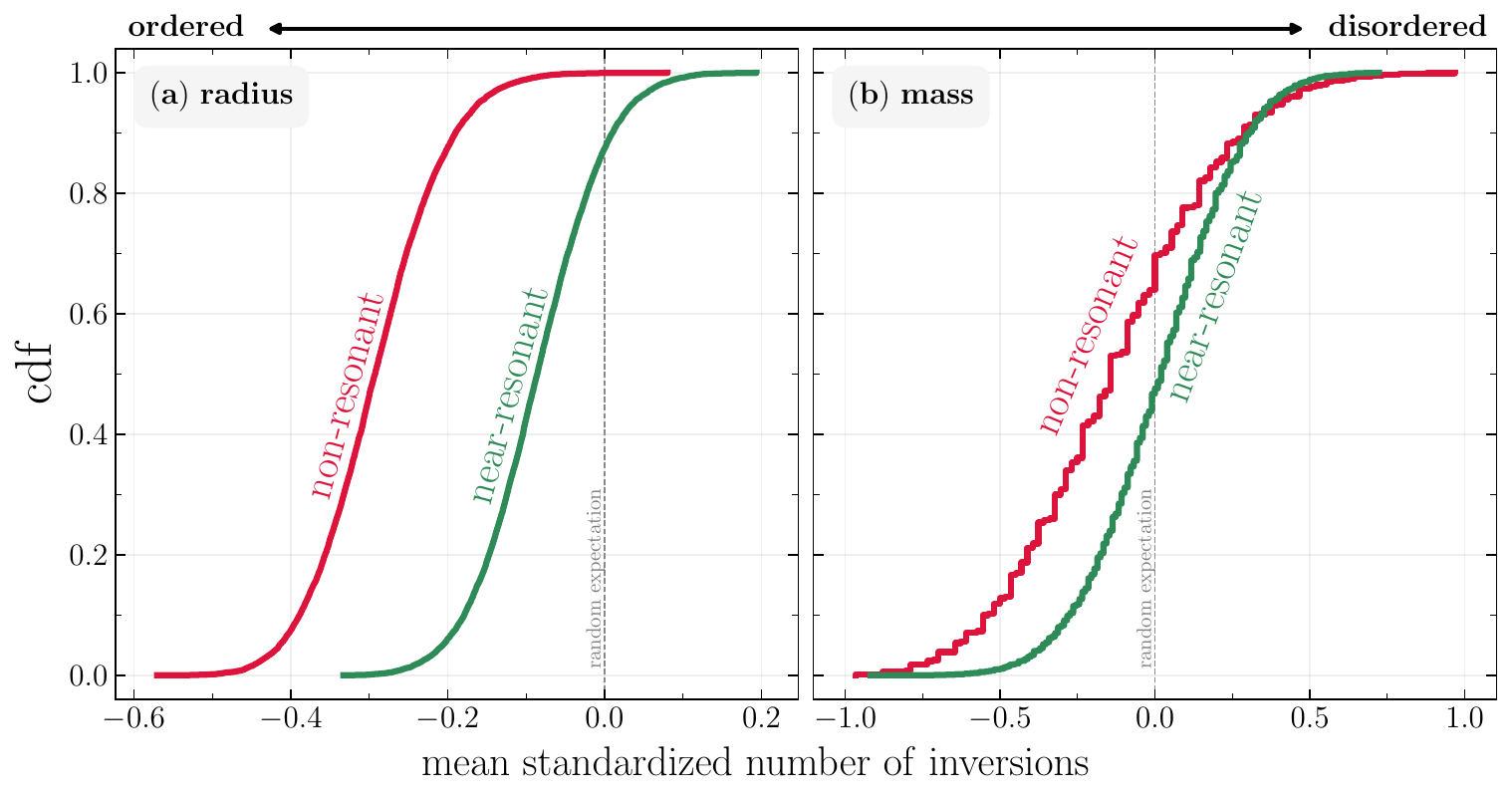}
    \caption{Comparison of planetary ordering between near-resonant and non-resonant systems. Cumulative distribution functions of the mean standardized ordering statistic, $Z_X$, for the two populations. Panel (a) uses the direction-independent number of inversions  with a radius threshold of 20\% for systems within the \textit{Kepler} sample, while panel (b) uses the same metric but for systems within the mass catalog. For each planetary system, the measured ordering statistic is standardized relative to the corresponding distribution obtained from all possible permutations of the sampled planetary properties at the same observed multiplicity, such that $Z_X=0$ (vertical dashed grey lines) represents the median expectation for random ordering, with increasingly negative values indicating progressively more ordered architectures. Near-resonant systems are systematically shifted toward higher values than non-resonant systems, indicating more disordered planetary radius and mass sequences.}
    \label{fig:main_result}
\end{figure*}

Because the near-resonant and non-resonant samples contain different proportions of three-, four-, and five-planet systems, we first compare their mean standardized ordering scores at fixed multiplicity $N$. We also account for finite sample uncertainty through bootstrap resampling. In particular, for each of the $10^4$ Monte Carlo realizations described in Section~\ref{sec:property_resampling}, we resample systems with replacement separately within each multiplicity $N$ and resonance class, preserving the original number of systems in each group. We then calculate the mean standardized ordering score for each resampled group. We show two of the resulting distributions in Figure \ref{fig:main_result}.

Using these mean standardized ordering scores, we calculate the contrast between the near-resonant and non-resonant populations at each multiplicity, allowing us to examine how their ordering difference varies with the number of planets:
\begin{equation}
\Delta Z_{X,N}=\bar{Z}_{X,N}^{(\mathrm{near\text{-}res})}-\bar{Z}_{X,N}^{(\mathrm{non\text{-}res})},
\label{eq}
\end{equation}
where $X$ denotes the adopted ordering metric, and $\bar{Z}_{X,N}$ represents the mean standardized ordering score for the corresponding population at multiplicity $N$.

Then, to combine these fixed-multiplicity contrasts into an overall population comparison, we assign each multiplicity class a weight based on the number of systems it contains across both populations. We define
\begin{equation}
w_N = \frac{n_N}{n_{\rm tot}},
\label{eq}
\end{equation}
where $n_N$ is the total number of systems with multiplicity $N$ across both populations, and $n_{\rm tot}$ is the total number of systems included in the comparison. The weights sum to unity and remain fixed across the Monte Carlo realizations. Using these weights, we combine the fixed-multiplicity contrasts to obtain the overall population contrast:
\begin{equation}
\Delta Z_X=\sum_{N=3}^5 w_N\,\Delta Z_{X,N}.
\label{eq}
\end{equation}

We summarize the resulting distributions of $\Delta Z_X$ using their 16th--84th percentile interval and $f_+$: the fraction of realizations for which $\Delta Z_X>0$. For the direction-independent metrics, positive contrasts in $\Delta Z_X$ indicate greater disorder among near-resonant systems, whereas negative contrasts indicate the opposite. Values of $f_+$ close to unity indicate that greater disorder among near-resonant systems is consistently recovered across realizations, whereas $f_+\approx0.5$ indicates no preferred direction. These summary statistics are presented in Figure~\ref{fig:radius_summary}.

\section{Results} \label{sec:results}

\subsection{Planetary Radius Ordering} \label{sec:radius_analysis}

The population comparison consistently favors less ordered radius sequences among near-resonant systems. Panel (a) in Figure~\ref{fig:main_result} illustrates this behavior for one representative analysis choice within the \textit{Kepler} sample. In particular, it shows the direction-independent number of inversions with a radius threshold of $20\%$: the near-resonant population is shifted toward higher values than the non-resonant population, corresponding to more disordered radius sequences. The bulk of each distribution nevertheless lies below the random-order reference, $Z_X=0$. Thus, near-resonant systems are not randomly ordered in an absolute sense; rather, they exhibit weaker ordering than their non-resonant counterparts. This distinction preserves the broader result that compact multi-planet systems possess non-random radius configurations \citep{Kipping2018}, while showing that the strength of this ordering is related to proximity to mean-motion resonances.

\begin{figure*}[t!]
    \centering
    \includegraphics[width=1\linewidth]{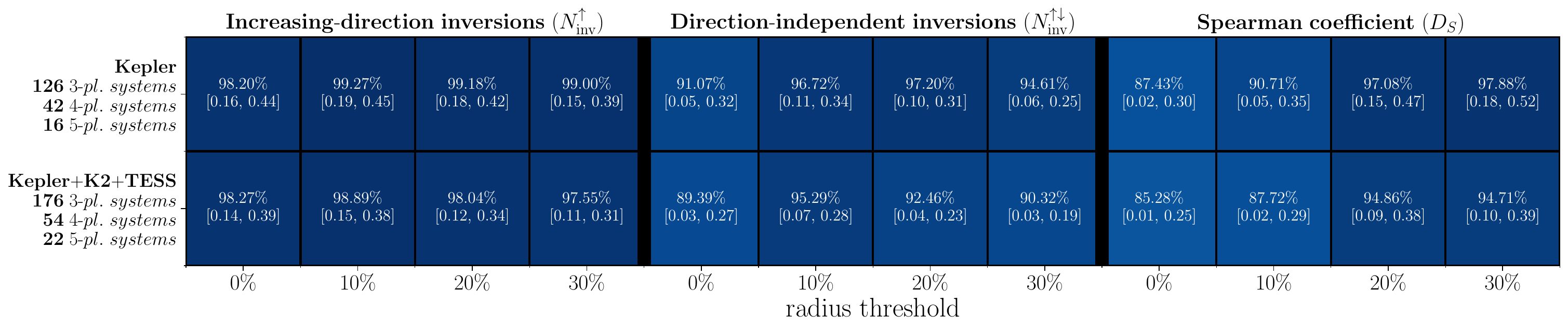}
    \caption{Summary of the near-resonant and non-resonant radius comparison. Each cell reports the fraction $f_+$ of the $10^4$ Monte Carlo resampling realizations for which $\Delta Z_X>0$. Values close to unity indicate that the near-resonant population is consistently more disordered across the resampling realizations. The bracketed values correspond to the 16th and 84th percentiles of the $\Delta Z_X$ distribution. Results are shown for the \textit{Kepler} and \textit{Kepler}+\textit{K2}+\textit{TESS} catalogs using the Spearman coefficient, the increasing-direction and the direction-independent number of inversions as measures of planetary ordering. All samples include confirmed planets and planet candidates and adopt the resonance classification criteria of Equations \ref{ec:delta_1st} and \ref{ec:delta_2nd}.}
    \label{fig:radius_summary}
\end{figure*}

We also quantify the magnitude and consistency of these population shifts using the distributions of the multiplicity-weighted contrasts $\Delta Z_X$, shown in Figure~\ref{fig:radius_summary}.

For the increasing-direction number of inversions, the \textit{Kepler} sample exhibits positive contrasts across all adopted radius thresholds, with $f_{+}$ ranging from $98.20\%$ to $99.27\%$ and the corresponding 16th--84th percentile intervals remaining above zero throughout. The combined \textit{K2T} (\textit{Kepler}+\textit{K2}+\textit{TESS}) sample recovers the same trend, with $f_{+}$ values between $97.55\%$ and $98.89\%$. The direction-independent inversion statistic yields the same qualitative result. In the \textit{Kepler} sample, $f_{+}$ ranges from $91.07\%$ to $97.20\%$, while the combined \textit{K2T} sample yields values between $89.39\%$ and $95.29\%$. For both samples, the 16th--84th percentile intervals remain positive across all thresholds. Together, the two inversion metrics indicate that near-resonant systems exhibit weaker monotonic radius ordering overall, rather than merely showing greater departures from outwardly increasing radius sequences.

The Spearman disorder statistic provides further support for this trend. In the \textit{Kepler} sample, $f_{+}$ increases from $87.43\%$ for the unfiltered sample to $97.88\%$ for systems with $\bar{\delta}_R>0.3$. The combined \textit{K2T} sample shows similar behavior, with $f_{+}$ increasing from $85.28\%$ to $94.71\%$ over the same thresholds. All corresponding 16th--84th percentile intervals remain above zero. Thus, the Spearman results reinforce the inversion-based findings, particularly for systems with larger mean pairwise radius contrasts. The agreement across these complementary metrics indicates that the observed trend is not specific to a particular measure of monotonic ordering.

Finally, the fixed-multiplicity comparisons presented in appendix Figure~\ref{fig:multiplicity_comparisons} show that the strongest and most consistent evidence comes from three-planet systems. For $N=3$, all three ordering metrics favor greater radius disorder among near-resonant systems across every adopted threshold, with all 16th--84th percentile intervals remaining above zero. This behavior is recovered in both the \textit{Kepler} and combined \textit{K2T} samples. In contrast, the four- and five-planet comparisons are less conclusive, with broader contrast distributions and less consistent trends across metrics and thresholds. These weaker constraints may reflect the substantially smaller sample sizes at higher multiplicities, rather than the absence of an underlying ordering difference.

\subsection{Planetary Mass Ordering} \label{sec:mass_analysis}

Planetary radii can be modified after formation by processes such as atmospheric loss and thermal evolution (e.g., \citealt{Lopez+2012, Owen+2013, Lopez+2013, Lopez+2014}). Thus, we examine whether the ordering difference identified in the radius analysis is also present in a more directly dynamical property: planetary mass. Nonetheless, and given the small sample size, we regard the mass comparison between near-resonant and non-resonant systems as a complementary consistency check rather than as an independent population-level measurement.

\begin{figure*}[t!]
    \centering
    \includegraphics[width=1\linewidth]{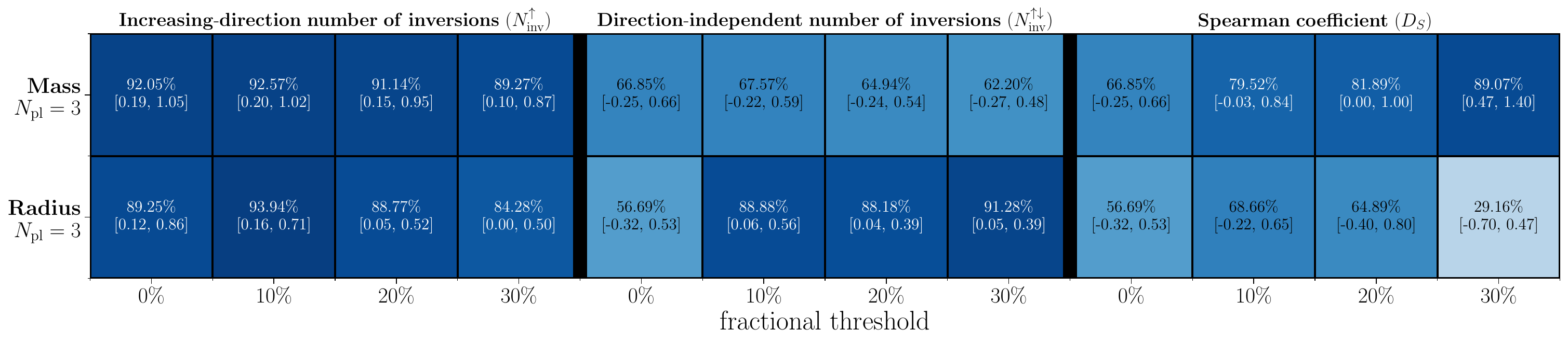}
    \caption{Summary of the near-resonant and non-resonant mass comparison. Each cell reports the fraction $f_+$ of the $10^4$ Monte Carlo resampling realizations for which $\Delta Z_X>0$. Values close to unity indicate that the near-resonant population is consistently more disordered across the resampling realizations. The bracketed values correspond to the 16th and 84th percentiles of the $\Delta Z_X$ distribution. Results are shown for the mass catalog using the Spearman coefficient, the increasing-direction and the direction-independent number of inversions as measures of planetary ordering. This sample only includes confirmed planets and adopts the resonance classification criteria of Equations \ref{ec:delta_1st} and \ref{ec:delta_2nd}. The corresponding radius results are included for reference.}
    \label{fig:mass_summary}
\end{figure*}

As shown by panel (b) in Figure~\ref{fig:main_result} and the summary table in Figure~\ref{fig:mass_summary}, the mass sample generally favors the same qualitative trend as the radius analysis: near-resonant systems tend to exhibit greater departures from monotonic mass ordering than non-resonant systems. This tendency is most consistently recovered by the increasing-direction inversion metric, while the direction-independent inversion metric shows weaker support. The Spearman disorder metric also favors the same trend, particularly at higher fractional thresholds. However, the mass comparison remains limited by the small sample of only 17 transiting three-planet systems. The corresponding radius results for this sample are included in Figure~\ref{fig:mass_summary} for reference.

\section{Discussion} \label{sec:discussion}

\subsection{Comparison with Previous Work} \label{sec:ordering_discussion}

Although our work is the first to systematically compare the monotonic ordering of entire planetary radius and mass sequences between near-resonant and non-resonant systems, earlier studies have explored different measures and associations of planetary radius ordering, including pairwise size trends near commensurabilities and correlations with orbital eccentricity.

\citet{Lozovsky+2025} investigated radius ordering
in \textit{Kepler} multi-planet systems and found a general preference for smaller inner planets. However, they found no statistically significant difference in the radius ratio distributions of neighboring pairs near and far from first-order period commensurabilities. Their comparison focused on individual planet pairs and included two-planet systems, whereas our analysis requires at least three planets and quantifies the monotonic ordering of entire planetary sequences. We also adopt different criteria for identifying proximity to mean-motion commensurabilities. Therefore, their result for pairwise radius ratios does not directly contradict our finding that near-resonant systems exhibit less monotonic radius ordering.

More recently, \citet{Gilbert+2026} investigated the relationship between radius monotonicity and orbital eccentricity. They found no statistically significant association, although systems with stronger outward-increasing radius trends tended to have higher eccentricities. Our finding that non-resonant systems exhibit stronger outwardly increasing radius ordering could be related to this tentative trend if these systems also have systematically higher eccentricities, although this remains to be established.

\subsection{Implications for Dynamical Evolution}
\label{sec:dynamical_interpretation}

The direction of the observed trend challenges the interpretation of proximity to resonance as a simple measure of how much primordial ordering information a system has retained. If planet formation and migration initially establish ordered planetary sequences that are subsequently disrupted by dynamical instabilities, non-resonant systems would be expected to exhibit greater radius and mass disorder than systems that remain near resonance. Instead, we find the opposite behavior: non-resonant systems exhibit more monotonic ordering than their near-resonant counterparts in both radius and mass. 

In the context of the \textit{breaking-the-chains} scenario, in which resonant chains assembled during disk-driven migration are subsequently disrupted by dynamical instabilities \citep{Izidoro+2017, Goldberg+2022}, one possible interpretation is that such instabilities preferentially preserve or produce more monotonically ordered planetary sequences among the surviving planets. Planet loss can change which planets remain in a system, while collisions and mergers can modify their masses and radii, potentially reshaping the ordering of both properties rather than simply randomizing an initially ordered sequence. Numerical studies have shown that dynamical instabilities can preserve or reshape intra-system mass uniformity and orbital spacing \citep{Goldberg+2022, Lammers+2023}. However, whether these processes also tend to enhance monotonic radius or mass ordering remains to be established.

Alternatively, near-resonant systems may more closely preserve primordial radius and mass correlations without those correlations necessarily corresponding to a preferred ordering with orbital period. The enhanced radius uniformity observed among near-resonant systems does not by itself imply stronger radius ordering \citep{Goyal+2023}. These possibilities are not mutually exclusive, and present observations cannot determine whether the ordering contrast was inherited at formation or produced by subsequent dynamical instabilities.

Planetary radius introduces an additional complication because it can evolve substantially after the orbital architecture has been established. Photoevaporation, core-powered mass loss, thermal contraction and tidal heating can alter planetary radii without necessarily rearranging the planets themselves \citep{Lopez+2012, Lopez+2013, Lopez+2014, Owen+2017, Ginzburg+2018, Millholland+2019a, Millholland+2019b}. In particular, obliquity-driven tidal heating may inflate the radii of short-period planets and has been linked to orbital configurations near mean-motion resonance \citep{Millholland+2019a, Millholland+2019b}. Because these processes depend on orbital period, planetary mass, composition and age, they can modify radius ordering independently of dynamical rearrangement. However, the qualitatively similar trend recovered in the mass sample suggests that radius evolution alone is unlikely to account for the observed ordering contrast. 

\subsection{Observational Limitations} \label{sec:limitations}

Several observational effects should be considered when interpreting our results. The observed architectures are mainly based on detected transiting planets, and undetected planets could alter both the measured multiplicity and the inferred radius sequence. However, our restriction to systems with at least three detected transiting planets already mitigates this effect, as multi-transiting systems are preferentially associated with dynamically compact and relatively coplanar architectures \citep{Lissauer+2011, Rowe+2014}. In addition, the high reliability of \textit{Kepler} multi-planet systems and their low false positive rates provide a robust foundation for population-level architectural studies \citep{Lissauer+2012, Rowe+2014}. Nevertheless, undetected planets cannot be completely excluded and could affect the observed multiplicity or introduce additional planets into the radius sequence.

Transit detectability also depends on planetary radius and orbital period. In particular, smaller planets are more difficult to detect at longer orbital periods, potentially biasing the observed sample toward larger outer planets and artificially enhancing outwardly increasing radius trends. Our comparisons at fixed observed multiplicity, combined with multiplicity-weighted contrasts, account for differences in the observed multiplicity distributions of near-resonant and non-resonant systems. Comparing the results from the \textit{Kepler} and \textit{Kepler}+\textit{K2}+\textit{TESS} samples further allows us to assess their sensitivity to survey composition. However, these procedures do not account for undetected planets that could bias the inferred radius ordering. Assessing the impact of observational selection effects on the measured ordering differences would require forward modeling of synthetic planetary populations through the survey detection process.

The mass sample is substantially smaller and less homogeneous than the radius sample, making the comparison more sensitive to small number statistics and selection biases associated with mass measurements. We therefore treat the mass ordering result as a complementary consistency check rather than an independent population-level constraint.

\section{Conclusions} \label{sec:conclusions}

We find that compact near-resonant multi-planet systems tend to exhibit weaker monotonic ordering in planetary radii and masses than their non-resonant counterparts. Using complementary rank- and inversion-based metrics, we recover this trend after accounting for differences in planet multiplicity and measurement uncertainties. The trend also persists in sensitivity tests that reduce the influence of planets with nearly equal radii or masses.

The direction of this trend is opposite to the simple expectation that disruption of initially ordered resonant chains should leave non-resonant systems more disordered. Together with the enhanced intra-system radius uniformity of near-resonant systems reported by \citet{Goyal+2023}, our findings highlight that uniformity and ordering describe distinct aspects of planetary architectures: systems can contain planets with similar radii or masses without exhibiting strong monotonic ordering with orbital period.

The evidence is strongest for planetary radii, while the smaller and less homogeneous mass sample provides a complementary consistency check. Present observations cannot determine whether the ordering difference was inherited during planet formation and migration, or produced by subsequent dynamical instabilities. Larger, more homogeneous samples with well-characterized detection completeness and improved planetary mass measurements, together with end-to-end simulations of planet formation and dynamical evolution, will be essential for understanding the origin of the observed trend.

\section*{Acknowledgments}

We thank Fred Adams, Gregory Gilbert, Fei Dai, Mu-Tian Wang, Christopher O'Connor and Hareesh Gautham Bhaskar for useful discussions and constructive feedback. S.W. was supported in part by the NASA Exoplanets Research Program NNH23ZDA001N-XRP (Grant No. 80NSSC24K0153), the NASA TESS General Investigator Program, Cycle 7, NNNH23ZDA001N-TESS (Grant No. 80NSSC25K7912), and the Heising-Simons Foundation (Grant No. 2023-4050). S.W. also acknowledges support through grant JWST-GO-09025.010-A, provided by NASA via the Space Telescope Science Institute under the JWST General Observers Program \#9025, and gratefully acknowledges support from the John and A-Lan Reynolds Faculty Research Fund. C.P. acknowledges support from the National Aeronautics and Space Administration under award No. 80NSSC26K1389 through the Exoplanets Research Program and support from the National Science Foundation under award No. AST-2511257.

This research has made use of the NASA Exoplanet Archive, which is operated by the California Institute of Technology under contract with NASA through the Exoplanet Exploration Program.

During the preparation of this manuscript, the authors used OpenAI's ChatGPT \citep{OpenAI+2026} (including GPT-5.6 Pro) to assist with manuscript organization and language editing, to generate alternative formulations, and to refine statistical notation, code presentation and figure design. The tool was not used to generate the observational data or reported numerical results. All AI-assisted text and code suggestions were critically reviewed, modified and validated by the authors. The authors verified all references, analyses and numerical results, and take full responsibility for the scientific content and conclusions of this work.

\software{NumPy \citep{Harris+2020}, SciPy \citep{Virtanen+2020}, Matplotlib \citep{Hunter+2007}.}

\renewcommand{\thefigure}{\thesection\arabic{figure}}
\setcounter{figure}{0}
\bibliographystyle{aasjournal}
\bibliography{sample631}

\appendix
\onecolumngrid

\section{Sample Composition by Multiplicity} \label{app:catalog_info}
\begin{table*}[h!]
    \centering
    \renewcommand{\arraystretch}{1.25}
    \begin{tabular*}{1\textwidth}
        {@{\extracolsep{\fill}}lccccccccc@{}}
        & \multicolumn{3}{c}{\textbf{Kepler}}
        & \multicolumn{3}{c}{\textbf{K2}}
        & \multicolumn{3}{c}{\textbf{TESS}} \\ [0.05cm]
        \cline{2-4}\cline{5-7}\cline{8-10} 
        &
        ${N_{\rm pl}=3}$ &
        ${N_{\rm pl}=4}$ &
        ${N_{\rm pl}=5}$ &
        ${N_{\rm pl}=3}$ &
        ${N_{\rm pl}=4}$ &
        ${N_{\rm pl}=5}$ &
        ${N_{\rm pl}=3}$ &
        ${N_{\rm pl}=4}$ &
        ${N_{\rm pl}=5}$ \\[0.1cm]
        \hline
        \hline
        \textbf{Non-resonant systems}
        & 70 & 20 & 3 
        & 15 & 3 & 0 
        & 12 & 0 & 0  \\ [0.1cm]
        \hline
        \textbf{Near-resonant systems}
        & 56 & 22 & 13 
        & 9 & 4 & 3 
        & 14 & 5 & 3  \\ [0.1cm]
        \hline
        \hline
    \end{tabular*}
    \caption{Number of near-resonant and non-resonant systems as a function of observed planet multiplicity for the \textit{Kepler}, \textit{K2}, and \textit{TESS} samples. Systems are classified as near-resonant if at least one neighboring pair satisfies the adopted first- or second-order resonance-proximity criterion. Systems with six or more planets are excluded because of the limited sample size.} 
    \label{tab:catalog_information}
\end{table*}

\section{Fixed-Multiplicity Population Comparisons} \label{app:multiplicity_results}

\begin{figure*}[h!]
    \centering
    \includegraphics[width=\linewidth]{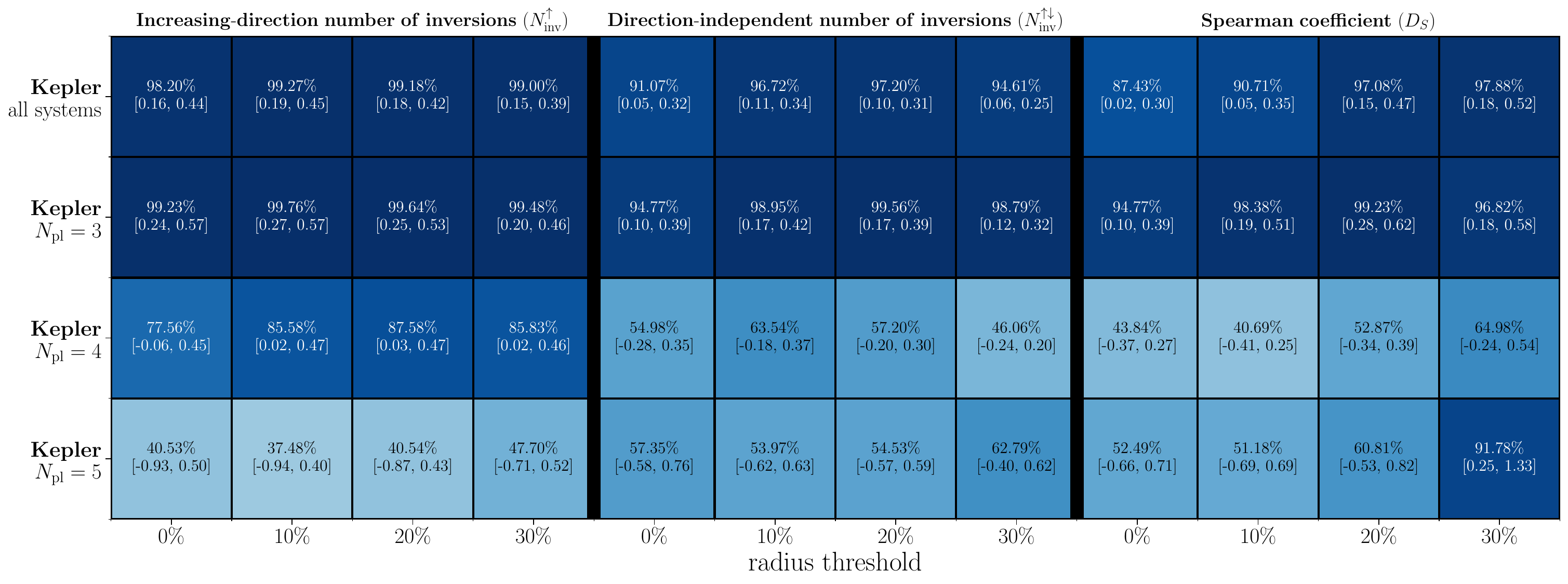}
    \vspace{1cm}
    \includegraphics[width=\linewidth]{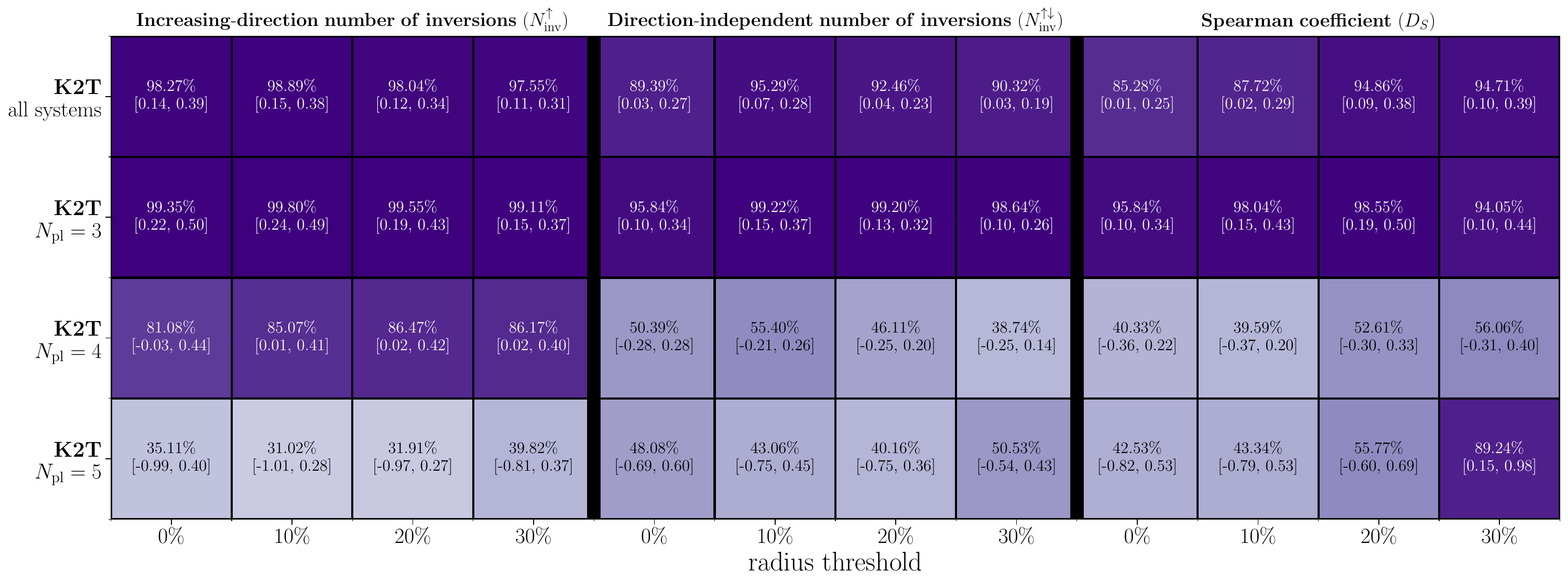}
    \caption{Fixed-multiplicity comparison of radius ordering between near-resonant and non-resonant systems. The top panel shows the \textit{Kepler} sample, while the bottom panel shows the combined \textit{Kepler}+\textit{K2}+\textit{TESS} sample. Each cell reports $f_+$, the fraction of the $10^4$ Monte Carlo realizations for which $\Delta Z_X>0$, with the 16th--84th percentile interval shown in brackets. Results are shown for the Spearman coefficient, the increasing-direction and direction-independent number of inversions. In each panel, the top row gives the multiplicity-weighted result, followed by the three-, four- and five-planet comparisons.}
    \label{fig:multiplicity_comparisons}
\end{figure*}

The figures below show the near-resonant and non-resonant comparison separately for three-, four- and five-planet systems, together with the multiplicity-weighted result for the full sample. These comparisons show how the individual multiplicity groups contribute to the overall trend and illustrate the weaker constraints at higher multiplicity.

Three-planet systems provide the strongest and most consistent constraints, as they dominate both samples. For these systems, all three ordering metrics favor greater radius disorder among near-resonant systems across all adopted thresholds, with the corresponding 16th–84th percentile intervals remaining above zero in both the \textit{Kepler} and combined \textit{Kepler}+\textit{K2}+\textit{TESS} samples. For four-planet systems, the increasing-direction inversion metric generally supports the same trend, particularly at nonzero radius thresholds, whereas the direction-independent inversion and Spearman metrics show no consistent preference between the two populations. The five-planet comparisons are substantially less conclusive: the increasing-direction inversion metric weakly favors the opposite trend, while the direction-independent metric shows no clear preference. The Spearman metric favors greater disorder among near-resonant systems only under the most restrictive selection criterion.

\end{document}